\documentclass[twocolumn, traditabstract]{aa} 
\usepackage{graphicx}
\usepackage{amsmath} 
\usepackage{subeqnarray}
\usepackage{layouts}
\usepackage{color}
    \definecolor{Blue}{rgb}{0.0,0.0,1.0}
    \definecolor{Red}{rgb}{1.0,0.0,0.0}
    \definecolor{Green}{rgb}{0.0,1.0,0.0}
\begin{document}
\title{Oscillations of the Eddington Capture Sphere}
%
\author{       Maciej Wielgus\inst{1, 2}
%
\and
               Adam Stahl\inst{3}
%
\and
               Marek Abramowicz\inst{3,1,4}
%
\and
               W{\l}odek Klu{\'z}niak\inst{1}
}
\institute{    Copernicus Astronomical Center, ul. Bartycka 18, PL-00-716
               Warszawa, Poland
\\ \email{wlodek@camk.edu.pl}
\and       Institute of Micromechanics and Photonics, ul. {\'s}w A. Boboli 8, PL-02-525, Warszawa, Poland
 \\ \email{maciek.wielgus@gmail.com}
\and          Physics Department, Gothenburg University,
               SE-412-96 G{\"o}teborg, Sweden
                 \\ \email{gusstaad@student.gu.se}
                 \\ \email{marek.abramowicz@physics.gu.se}
\and           Institute of Physics, Faculty of Philosophy and Science,
               Silesian University in Opava,
               Bezru{\v c}ovo n{\'a}m. 13, CZ-746-01 Opava,
               Czech Republic
}
   \date{Received Aug 14, 2012; accepted ????}
\abstract {We present a toy model of mildly super-Eddington, optically thin accretion onto a compact star in the Schwarzschild metric, which predicts periodic variations of luminosity when matter is supplied to the system at a constant accretion rate. These are related to the periodic appearance and disappearance of the Eddington Capture Sphere. In the model the frequency is found to vary inversely with the luminosity. If the input accretion rate varies (strictly) periodically, the luminosity variation is quasi-periodic, and the quality factor is inversely proportional to the relative amplitude of mass accretion fluctuations, with its largest value $Q\approx 1/(10\, |\delta \dot M/\dot M|)$ attained in oscillations at about 1 to 2 kHz frequencies for a $2M_\odot$ star.
}
\maketitle

\section{Introduction}
\cite{Abramowicz} (hereafter AEL) argued that the luminosity of a
relativistic star that is accreting at a super-Eddington rate, should
periodically change. They have shown that in the combined
gravitational and radiation fields of a~spherical, compact star,
radially moving test particles are captured by a sphere on which the
gravitational and radiative forces balance. This is because in
Einstein's general relativity the radiative force diminishes more
strongly with the distance than the gravitational force, and radiation
may be super-Eddington close to the star, but sub-Eddington further
away, reaching the Eddington value at the 
\emph{Eddington Capture Sphere} (ECS), whose radius is given
by a simple expression in the Schwarzschild coordinates (\cite{Phinney}),
\begin{equation}
\label{Eddington-radius}
r_{\rm ECS} =  \frac{2 R_G}{1 - \left(1 - \dfrac{2 R_G}{R^2} \right)^2
\left(\dfrac{L}{L_{\rm Edd}} \right)^2}\ .
\end{equation}
Here $R$ is the radius of the star, $R_G = GM/c^2$ its
gravitational radius, and $L$, which is assumed to satisfy
 Eq.~(\ref{cond}), is the stellar luminosity at its surface.  
\cite{Bini,Oh}, and \cite{Stahl} have shown
that the ECS captures particles from a wide class of non-radial orbits
as well\footnote{As shown in detail by \cite{Oh}, azimuthal radiation
  drag efficiently removes particle's angular momentum, typically
  making motions in the combined gravitational and radiation fields
  asymptotically radial.}, so the following discussion is not
restricted to the case of radial accretion. However, it is important
to keep in mind that the balance of forces necessary for the existence
of the ECS requires a large, radial radiative flux. In non-spherical
accretion this can only be realized in the optically thin regime.

 In this paper we present a simple model in which the idea
of periodic luminosity changes is realized.
AEL assumed that the radiation power was provided by the kinetic
 energy of the accreted particles, released when they hit the surface
 of the star.  As stressed by \cite{Stahl}, particles settle on the
 ECS rather gently, so the ECS itself is not particularly luminous.
 If all particles are captured at the Eddington sphere, they do not
 reach the surface of the star, and the stellar accretion luminosity
 goes to zero. When it does, and actually already when
 ${L}/{L_{\rm Edd}} < (1-2R_G/R)^{-1/2}$, the ECS disappears,
 accretion is resumed by the star and eventually
the luminosity may build up to its
former value, so the ECS will reappear and accretion will stop again.
Thus, the accretion process may be quasiperiodic,
alternating between states of high luminosity and no luminosity.

%
\section{The model}
The toy model considered here assumes a steady\footnote{The
assumption of a constant $\dot M$ will be relaxed in
Section~\ref{discuss}.} supply of optically thin fluid at some
  distance above the stellar surface. As shown by \cite{Stahl}, unless
  its velocity is extraordinarily large the fluid will settle on the
  ECS, regardless of its point of origin, if only
\begin{equation}
 (1-2R_G/R)^{-1/2}<{L}/{L_{\rm Edd}} < (1-2R_G/R)^{-1}.
\label{cond}
\end{equation}

In our model the stellar luminosity will be assumed to be either zero,
or to have a certain definite value, $L=L_0$, satisfying  Eq.~(\ref{cond}).
  In this sense the model states are binary (on-off). 
This property implies that the ECS is located at
a specific radius, $r_{ECS}=r_0>R$, whenever present 
(Eq.~[\ref{Eddington-radius}]). The model is described by three equations:
\begin{eqnarray}
\quad\quad L_{\rm ECS}(t) &=& \mu L_\ast(t - \delta t_l), \label{model-ECS}\\
L_\ast(t) &=& \eta {\dot M}_\ast(t -  \delta t_r), \label{model-surface} \\
{\dot M}_\ast(t) &=&
\begin{cases}{\dot M}_0 &\text{if}\ \   L_{\rm ECS}(t -  \delta t_i) <\mu L_0,\\
0 & \text{if}~ ~  L_{\rm ECS}(t -  \delta t_i) \ge \mu L_0\ ,
\end{cases} 
\label{model-accretion}
\end{eqnarray}
where $L_{ECS}$ is the stellar luminosity at the ECS, $L_\ast$ is the
luminosity at the stellar surface, $\dot{M}_\ast$ is the accretion
rate at the stellar surface, $\mu$ is a redshift factor between $R$ and
$r_0$, $\eta$ is the conversion factor between accretion rate and
stellar luminosity.

The simplest possibility, which we will ignore, is that
$\eta {\dot M}_0 <L_{\rm Edd}(1-2R_G/R)^{-1/2}$, so
the critical luminosity for establishing the Eddington
capture sphere is never reached, and the equations describe
a steady state solution. 
We will assume instead that $L_0=~\eta\dot M_0$, i.e., the stellar luminosity
in the ``on'' state has the necessary value to establish
an Eddington Capture Sphere at the radius $r_0$.
This assumption allows non-trivial time behaviour of the model system.

There are three timescales that determine the behaviour of the model,
the \emph{reaction time} $\delta t_r$, which is the timescale for
converting accreting matter into radiation, the 
\emph{light travel  time} $\delta t_l$, 
which is the travel time of light from the stellar surface to the
 Eddington Capture Sphere, and the \emph{infall  time} $\delta t_i$, 
which is the infall time of matter from the ECS
to the stellar surface in the absence of radiation. Now, $\delta
t_l$ and $\delta t_i $ can be determined from $R_G, R$ and $r_0$. So, for a
given star these two timescales are a function of the stellar
luminosity alone,  i.e., of $\eta\dot M_0$. The reaction time
introduces a phase shift between accretion and radiation at the
stellar surface. In numerical examples we will adopt two extreme
values $\delta t_r=0$, or $\delta t_r=0.1\,$ms, the latter being the
estimated cooling time of a clump of fluid that falls on a neutron
star surface (\cite{KMW}).

In reality, the time delays would not be as sharp as we have assumed them
to be. For instance, even at time $\delta t_l$ after the luminosity is
turned off, particles outside the ECS suffer from radiation
drag for an additional interval of time $(r-r_0)/c$, where $r$ is the
position of the particle at time $\delta t_l+(r-r_0)/c$ after the radiation
is turned off. Conversely, after the luminosity is turned on again, all
particles between the ECS and the stellar surface will feel the full impact
of radiation pressure after a delay of only $(r-R)/c<\delta t_l$,
where again, $r$ is the position of the particle when the radiation front
passes it, $(r-R)/c$ after the radiation is turned on. In the toy
model, we neglect such effects entirely.

Combining Eqs.~(\ref{model-ECS})-(\ref{model-accretion}) 
we find that $\dot{M}_\ast(t)$ is
determined by $\dot{M}_\ast(t-T)$, with $T=\delta t_i + \delta t_r +\delta t_l$:

\begin{figure}
\begin{center}
\includegraphics[width=0.50\textwidth]{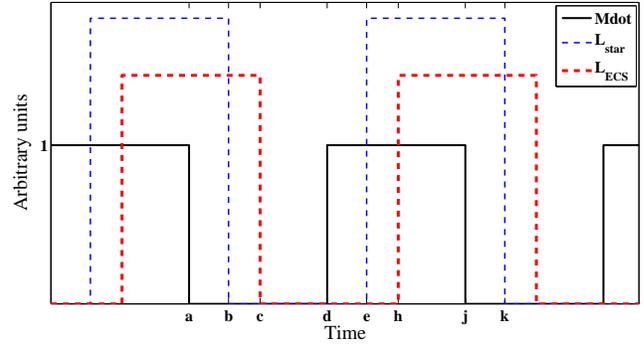}
\end{center}
\caption{Stellar accretion rate and luminosities at the stellar surface
and at the Eddington Capture Sphere, according to the model. }
 \label{fig:behaviour}
\end{figure}
\begin{eqnarray}
\dot{M}_\ast(t)
 &=&{\dot M}_0\, \Theta\left(\mu L_0 - L_{\rm ECS}(t -  \delta t_i)\right)
\nonumber\\
 &=&{\dot M}_0\,  \Theta\left( L_0 - L_\ast(t - \delta t_i - \delta t_l)\right)
\nonumber\\
 &=&{\dot M}_0\, 
 \Theta\left({\dot M}_0 -\dot{M}_\ast(t-\delta t_i -\delta t_l -\delta t_r)
 \right)\nonumber\\
 &=&{\dot M}_0 - \dot{M}_\ast(t-T),
\nonumber
\end{eqnarray}
%
%
%
where $\Theta$ is the Heaviside step function.
Thus,
\begin{eqnarray}
\left\{ {\dot M}_\ast (t - T) = {\dot M}_0\right\}
   &\Rightarrow& \left\{ \dot{M}_\ast(t) = 0 \right\}, \nonumber \\
\left\{ {\dot M}_\ast (t - T) = 0\right\}
 &\Rightarrow&  \left\{ \dot{M}_\ast(t) = {\dot
M}_0 \right\},
 \nonumber
\label{implication-double-period}
\end{eqnarray}
and clearly, the model system shows periodic behaviour with the period $2T$:
\begin{equation}
\dot{M}_\ast(t) = \dot{M}_\ast(t- 2T)\ . \label{eq:periodicity}
\end{equation}

\begin{figure}
\begin{center}
\includegraphics[width=0.50\textwidth]{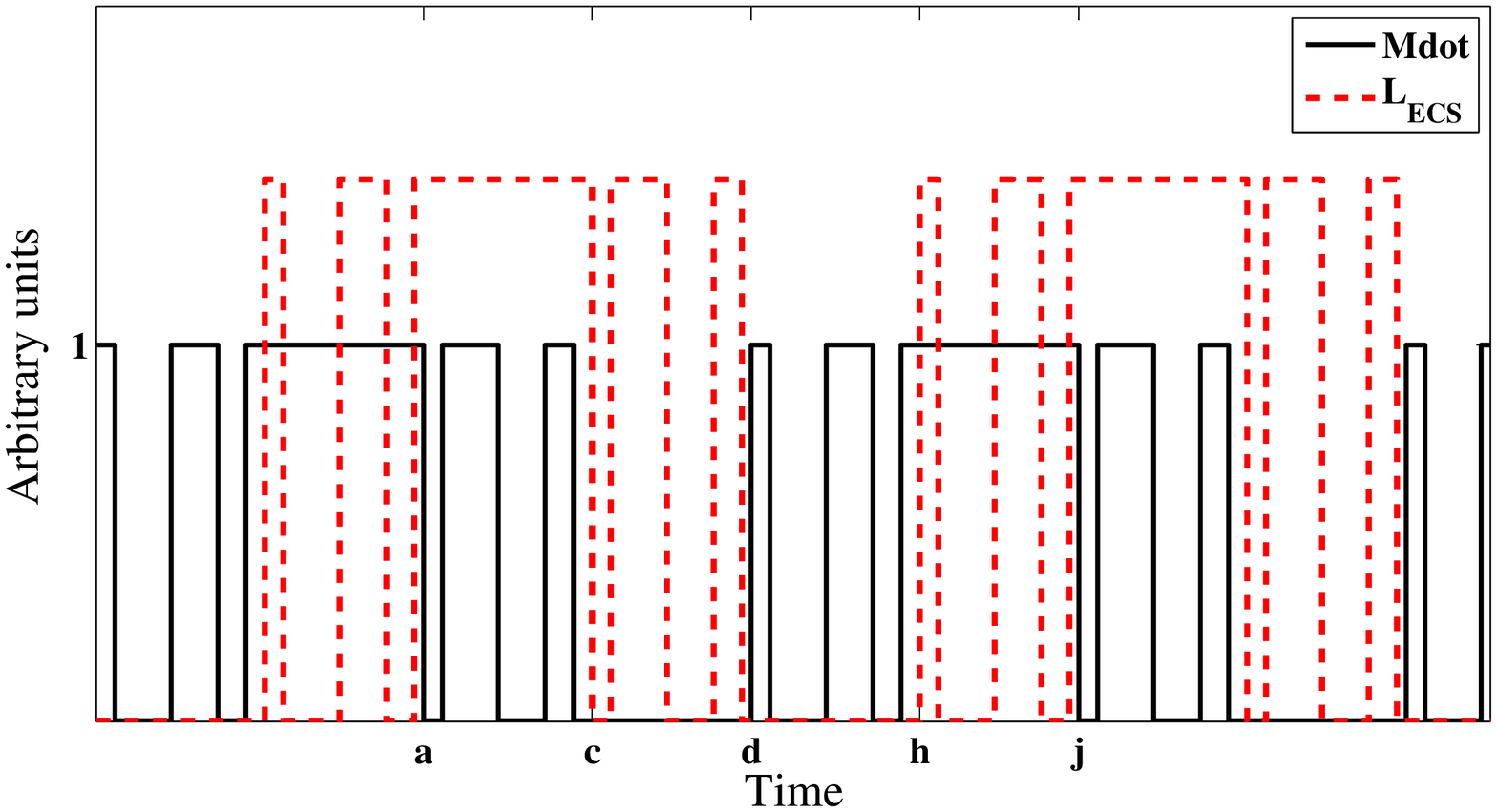}
\end{center}
\caption{Stellar accretion rate and the luminosity at the ECS,
for another choice of initial conditions. }
 \label{saw}
\end{figure}
\begin{figure}
\begin{center}
\includegraphics[width=0.50\textwidth]{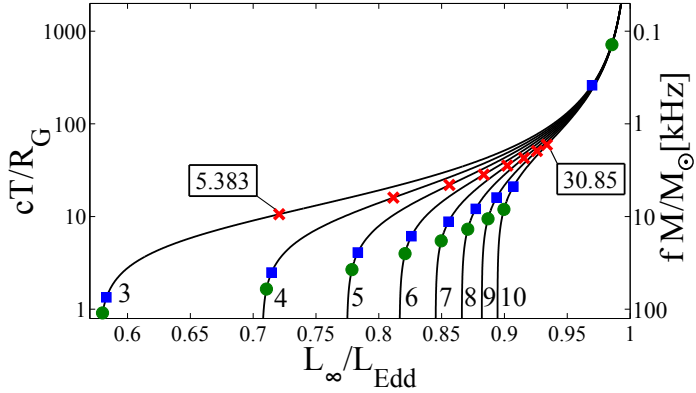}
\end{center}
\caption{The semi-period $T$ of the oscillation in geometrical units
as a function of peak luminosity at infinity in Eddington units,
when kinetic energy is instantaneously converted to luminosity
($\delta t_r=0$).
The corresponding frequency $f=1/(2T)$ can be read off in kHz
from the right vertical axis.  The curves are labeled with
the stellar radius (3,..,10) in units of $R_G$.
Filled squares indicate a value of the slope
$d\log T/d\log  L_\infty=50$, filled circles a value of 100,
and the crosses the minimum value of the logarithmic derivative
for each curve.
}
 \label{fig:relation}
\end{figure}
\section{Results of the model}

A typical behaviour of the model is shown in Fig.~\ref{fig:behaviour}
which shows a solution of
Eqs.~(\ref{model-ECS})-(\ref{model-accretion}) with  
$\delta t_l :\delta t_i :\delta t_r=8:17:10$. 
The black solid line traces ${\dot M}_\ast$.
In the figure, we identify the time $b-a=e-d =\delta  t_r$ as the reaction
time, $c-b=h-e =\delta  t_l$ as the light travel time, and $d-c =j-h=\delta
t_i$ as the infall time.  The luminosities at the stellar surface and
at the ECS are shifted relative to the accretion rate by $\delta t_r$
and  $\delta t_r + \delta t_l$, respectively. We note also that 
$d-a =j-d=h-c=k-e = e-b = \delta t_r + \delta t_l + \delta t_i = T$,
 and the pulses are as
long as the periods without activity. Thus the period of the
oscillation is $P = 2T$, as expected from Eq \eqref{eq:periodicity}.

In constructing Fig.~\ref{fig:behaviour} it was assumed that matter
first arrives at the stellar surface at $t=0$ and that there is a
continual inflow of matter to the system at the rate $\dot M_0$ for
all $t\ge0$.   Different initial conditions can lead to a more
complicated pulse shape of ${\dot M}_\ast(t)$ for $t$ in the intervals
$(2nT,2nT+~T)$, with its complement 
${\dot M}_\ast(t)=\dot M_0-{\dot M}_\ast(t-T)$ in
the intervals $(2nT+T,2nT+2T)$. Here, and elsewhere, $n=0,1,2,3...$.
  The state of accretion ($\dot M_0$ or
0) at time $t$ has no influence on the future value of ${\dot M}_\ast$
until the time $t+T$, so one can assume that at any instant in the
initial interval $t\in[0,T)$ the accretion rate has any of the two
  values, 0 or $\dot M_0$, i.e., in this interval 
${\dot M}_\ast(t)=\dot M_0g(t)$, where $g(t)$ is an arbitrary binary
  function, mapping the initial time interval into on-off states,
  $g:[0,T)\rightarrow\{0,1\}$.  Fig.~\ref{saw} provides an example of
    such behaviour. Thus, in principle,
the harmonic content of the signal may
    be quite rich, although the fundamental is still at $f=1/P=1/(2T)$.

As can be seen from
Eqs.~ (\ref{model-ECS})-(\ref{model-accretion}) the model takes no
account of accumulation of matter on the ECS, that is to say, here we
ignore the arrival at $t=2(n+1)T$
 (and at other moments as well in the case of Fig.~\ref{saw}),
of the additional matter that had accumulated on the ECS.
In reality, this matter would produce
a brief flash of radiation of energy $\eta_{\rm ECS}\dot M_0\Delta T$, where
$\eta_{\rm ECS}$ is the conversion efficiency to radiation
of the kinetic energy of matter falling from the ECS, and $\Delta T$ is the
accumulation time ($\Delta T=T$ in  Fig.~\ref{fig:behaviour}).
If in the steady accretion phase matter is 
falling in from $r>>r_0$, one would expect $\eta_{\rm ECS}<<\eta$.

\begin{figure}
\begin{center}
\includegraphics[width=0.50\textwidth]{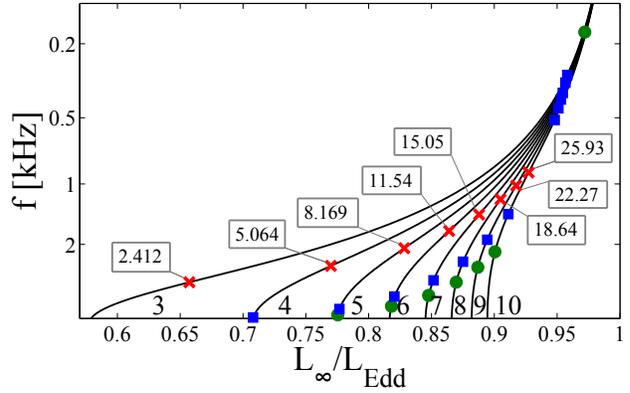}
\end{center}
\caption{The frequency of the oscillations
as a function of peak luminosity at infinity in Eddington units,
when there is a delay in converting kinetic energy to luminosity
$\delta t_r=0.1\,$ms (see \cite{KMW}).
 The stellar mass is assumed to be $M=2M_\odot$.
The curves are labeled with
the stellar radius (3,..,10) in units of $R_G$.
Filled  squares indicate a value of the slope
$d\log T/d\log  L_\infty=30$, filled circles a value of 50,
and the crosses the minimum value of the slope
for each curve (boxed values).
}
 \label{fig:mass}
\end{figure}

In a physically realistic situation  the infall time
 is the dominant time scale, $\delta t_i \gg \delta t_l$. The radius of the
Eddington Capture Sphere, and hence also the infall time, increases rapidly
with $L_\infty$, and without bounds as $L_\infty\rightarrow L_{\rm Edd}$:
\begin{equation}
\label{ECS}
\frac{r_{\rm ECS}}{2R_G} = {\left[1 -
\left(\dfrac{L_\infty}{L_{\rm Edd}} \right)^2\right]^{-1}}\ .
\end{equation}
Here, $L_\infty$ is the redshifted luminosity
at infinity.
Hence, except for the lowest values  of the luminosity parameter $L_0$,
the infall time $\delta t_i$ is the dominant time scale, and the
frequency of oscillations $f=1/(2T)$ varies inversely with the luminosity.
Neglecting $\delta t_r$, we can compute the semi-period $T$, and the frequency of
oscillation as a function of the stellar radius $R$ and of the luminosity
$L_\infty$ alone---$T$ and $f$ will, respectively, scale directly or inversely with $M$. These are shown in Fig.~\ref{fig:relation} for various stellar radii.
Fig.~\ref{fig:mass} shows the frequency as a function of luminosity,
for a star with $M=2M_\odot$, when $\delta t_r=0.1\,$ms.

\section{Conclusions and discussion}
\label{discuss}

We have shown that supplying mass to the vicinity of a compact star,
continuously and at a constant accretion rate, can lead to  periodic top-hat
variations of luminosity, if only the mass accretion rate corresponds to
a mildly super-Eddington luminosity at the stellar surface.
This oscillatory behaviour of luminosity is related to the phenomenon
of the Eddington Capture Sphere, which is a consequence
of  the interplay of radiation
drag and general relativity. 

If such an oscillation occurs in the real
world, for example in the Z sources, where rapid variations of the inferred
inner radius of the accretion disk have been reported (\cite{Lin}),
the actual mechanism is likely to be more complex than the strictly
periodic oscillation in the toy model considered here. 
E.g., the accretion rate
is not likely to be constant (contrary to our assumption of
simple on-off behaviour at the stellar surface).

As a minor extension of the model, consider an accretion rate that
is varying on a timescale comparable to the period of the ECS oscillator.
Even strictly periodic variation of $\dot M_0$ will lead to a decoherence
of the ECS oscillation, because the position $r_0$ of the ECS and
(hence) the infall time from the ECS, and (hence) the
oscillator frequency, are strongly varying functions of the luminosity,
$L_0=\eta M_0$. Indeed,
\begin{equation}
\label{decohere}
\frac{|\delta f|}{f} = \frac{|\delta T|}{T} 
 = \frac{d\ln T}{d\ln L_\infty}\frac{|\delta  L_\infty|}{ L_\infty}
  =\frac{d\ln T}{d\ln L_\infty}\left|\frac{\delta \dot M_0}{\dot M_0}\right|.
\end{equation}
Thus, the $Q$ factor of the oscillation is
\begin{equation}
\label{Q}
Q=\frac{f}{|\delta f|} = \left|\frac{\dot M_0}{\delta \dot M_0}\right|
  \left(\frac{d\ln T}{d\ln L_\infty}\right)^{-1}.
\end{equation}
We have indicated some values of the logarithmic derivative
${d\ln T}/{d\ln L_\infty}$ in Figs.~\ref{fig:relation},~\ref{fig:mass}. 
Comparing the figures, we see that
delayed emission (reaction time $\delta t_r>0$) 
at the stellar surface acts as a low-pass filter, cutting out
the highest frequencies. At the same time it improves the quality factor of
the oscillator.

Accreting neutrons stars in low mass X-ray binaries are expected to have
a radius in the range of
$4$ to $7 GM/c^2$ (\cite{arnett,kw}), and a mass of about two solar masses.
It is seen from Fig.~\ref{fig:mass} that the minimum values
of the logarithmic derivative attained  
for radii of 4, 5, 6, $7 GM/c^2$ are between 5 and 15.
Hence, the maximum expected value of the oscillator quality factor would be
$Q\approx 1/(10 |\delta\dot M_0/\dot M_0|)$, occurring for frequencies
between about 1 and 2 kHz,
as can be seen from Fig.~\ref{fig:mass}.

\section{Acknowledgements}

Research supported in part by Polish NCN grants
UMO-2011/01/B/ST9/05439 and N N203 511238. 


\end{document}